\documentclass[final,3p,times,twocolumn]{elsarticle}

\usepackage{epsfig}

\usepackage{amssymb}
\usepackage{lineno}
\usepackage{upgreek}
\usepackage{subfigure}
\usepackage{pifont}
\usepackage{color}

\biboptions{sort&compress}

\journal{NIM A}

\begin{document}

\begin{frontmatter}



\title{Comparison of Modeled and Measured Performance of a GSO Crystal as Gamma Detector}
\date{}

\author[cmu,cenpa]{D.\,S.~Parno\corref{cor1}}
\author[cmu,kek]{M.~Friend}
\author[cmu]{V.~Mamyan}
\author[cmu]{F.~Benmokhtar\fnref{FBpresaddress}}
\author[jlab]{A.~Camsonne}
\author[cmu]{G.\,B.~Franklin}
\author[uva]{K.~Paschke}
\author[cmu]{B.~Quinn}

\fntext[FBpresaddress]{Present address: Duquesne University, Pittsburgh, PA 15282, USA}
\cortext[cor1]{Corresponding author. dparno@uw.edu}

\address[cmu]{Carnegie Mellon University, Department of Physics, Pittsburgh, PA 15213, USA}
\address[cenpa]{University of Washington, Center for Experimental Nuclear Physics and Astrophysics and Department of Physics, Seattle, WA 98195, USA}
\address[kek]{High Energy Accelerator Research Organization (KEK), Tsukuba, Ibaraki, Japan}
\address[jlab]{Thomas Jefferson National Accelerator Facility, Newport News, VA 23606, USA}
\address[uva]{University of Virginia, Department of Physics, Charlottesville, VA 22904, USA}

\begin{abstract}

We have modeled, tested, and installed a large, cerium-activated Gd$_2$SiO$_5$ crystal scintillator for use as a detector of gamma rays. We present the measured detector response to two types of incident photons: nearly monochromatic photons up to 40~MeV, and photons from a continuous Compton backscattering spectrum up to 200~MeV. Our GEANT4 simulations, developed to determine the analyzing power of the Compton polarimeter in Hall A of Jefferson Lab, reproduce the measured spectra well.

\end{abstract}


\end{frontmatter}

\section{Introduction} \label{sec:intro}

Cerium-activated gadolinium oxyorthosilicate (Gd$_2$SiO$_5$:Ce, or GSO)~\cite{Takagi:CeDopedGSO1983} is a scintillator with a high light output and fast decay time relative to many other commonly used scintillating crystals, \textit{e.g}.\ Bi$_4$Ge$_3$O$_{12}$ (BGO). These properties, along with the crystal's non-hygroscopic nature, its relative ease of growth, and its radiation hardness, have made it a popular choice for a number of detection applications. Its most high-profile use is in positron emission tomography~\cite{Humm:PETreview_2003}, where detectors are optimized for 511-keV photons, but GSO scintillators have also been used to detect protons~\cite{Tamagawa:ConnectedGSOProton_2006}, charged leptons, and pions~\cite{Kobayashi:GSO_EM_cal_1991}.

In 2009, a GSO crystal with Ce:0.5~mol\% doping, grown by Hitachi Chemical and read out with a photomultiplier tube (PMT), was adopted as a gamma detector for the upgraded Compton polarimeter~\cite{Friend:ComptonUpgrade2011} in Hall A~\cite{Alcorn:HallA04} of the Thomas Jefferson National Accelerator Facility (Jefferson Lab)~\cite{Leemann:CEBAF01}. This device exploits Compton scattering to make a continuous measurement of the longitudinal electron-beam polarization, a vital parameter for a significant portion of Hall A's experimental program. Integration of the energy that backscattered photons deposit in this crystal allows such a measurement to be made with precision better than 1\%~\cite{Friend:ComptonUpgrade2011}. With a Compton-laser wavelength of 1064~nm, as in the Jefferson Lab data described in this work,  the maximum energy of a Compton-backscattered photon may range from 18~to~580~MeV, depending on the beam energy chosen for the experiment ($1-6$~GeV). 

Uncertainties in the detector response to incident gammas are important potential sources of systematic error. In particular, the energy-weighted integration is sensitive to non-linearities in the response. To reduce these uncertainties, a model was developed using the GEANT4 simulation toolkit~\cite{Agostinelli:Geant4_03} and was compared to calibration data, taken at two facilities, with incident gammas from 20 to 200~MeV. Section~\ref{sec:sim_comptspec} describes the Compton polarimeter and the fundamental simulation method as applied to single-arm Compton photon data at Jefferson Lab. Section~\ref{sec:jlab} discusses further tests at Jefferson Lab using photons tagged by coincident Compton-scattered electrons. In Section~\ref{sec:higs}, we give the results of tests with nearly monoenergetic photon beams in the $20-40$-MeV range at the High-Intensity $\upgamma$ Source (HI$\upgamma$S)~\cite{Weller:HIGS09}, a facility on the Duke University campus that produces gammas by Compton backscattering of light stored in a free-electron laser cavity. 

\section{Hall A Compton Polarimeter and Simulation Method} \label{sec:sim_comptspec}

\begin{figure}[] 
\begin{center}
\includegraphics[width=\columnwidth]{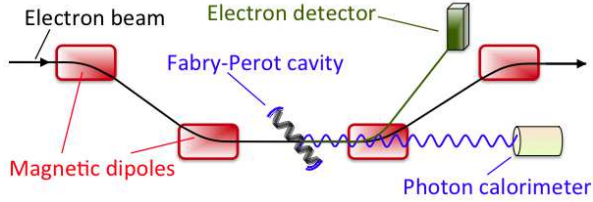} 
\caption[]{Schematic of the Hall A Compton polarimeter~\cite{Friend:ComptonUpgrade2011}.}
\label{fig:compton_schematic} 
\end{center} 
\end{figure}

In the Compton polarimeter in the Hall A beamline at Jefferson Lab, the polarized electron beam is routed through a four-dipole magnetic chicane (Fig.~\ref{fig:compton_schematic}). In the center of the chicane, the beam interacts with circularly polarized laser light in a high-finesse Fabry-P\'erot cavity. Compton-backscattered photons pass undeflected through the one-inch gap~\cite{Bardin:ComptonCDR96} in the third dipole; the GSO calorimeter is located on this direct path. A silicon-microstrip detector (Section~\ref{sec:jlab}) detects Compton-scattered electrons, which are deflected through a larger angle than the unscattered majority of the beam. The Fabry-P\'erot cavity is periodically taken out of resonance, nearly eliminating Compton-scattering events in the cavity and allowing a direct background measurement. During typical running, the dipole fields are controlled via feedback from a beam-position monitor in the chicane, so as to maintain a stable beam position.

Light from the GSO calorimeter is collected in a 2-inch, 12-stage BURLE Industries RCA~8575 PMT, with a base customized for maximum linearity of response. The data-ac\-quisition system is based on a 12-bit Struck SIS3320 flash analog-to-digital converter, modified to integrate the input signal over an externally timed window and configured to sample at 200~MHz. In addition to performing this onboard integration, the SIS3320 card records all samples from the window in one of two internal buffers. Time\-stamps from photon- or electron-detector triggers, recorded in a CAEN V830 latching scaler, allow the retention of some information about individual pulses. For a prescaled sample of pulses, the numerical sum of samples from the programmable readout window is recorded in the datastream; the energy of the incident photon can be retrieved from this pulse integral. For a smaller number of pulses, all of the samples for the readout window are written to disk, allowing pulse-shape analysis~\cite{Friend:ComptonUpgrade2011}. It is also possible to trigger the system at regular intervals that are uncorrelated with photon pulses, allowing the study of pileup events. These triggers are generated either in software or with a remotely programmable function generator.

\begin{figure}[] 
\begin{center}
\includegraphics[width=\columnwidth]{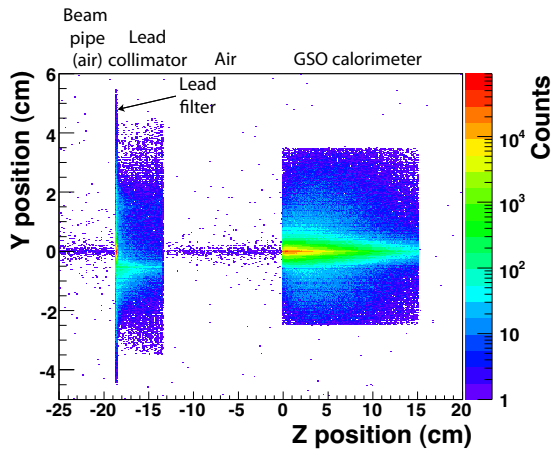} 
\caption[]{Histogram of positions of the first interaction point of incident photons in the Hall A Compton gamma beamline, as determined by the GEANT4 MC. The $z$ position is measured along the central axis of the incident Compton-scattered photons; the $y$ axis is vertical. In this figure, the central axes of the calorimeter and of the 2-cm aperture of the collimator are offset by 0.5~cm from the gamma beam, reflecting conditions during one run period.}
\label{fig:posfirstEdep} 
\end{center} 
\end{figure}

The GEANT4 Monte Carlo (MC) simulation of the GSO crystal response was performed with version 4.9.4, patch~03. The MC begins with the generation of a beam of simulated photons to match the experimental beam. To reproduce a Compton-backscattered beam, simulated photons of various energies are generated with probabilities weighted by the Compton scattering cross-section for the specific initial electron and laser-photon energies of the planned experiment. No other electron-beam properties are included in the model. The simulated photons are then allowed to interact with beamline items downstream of the dipole; in the standard Hall A installation, this includes a 0.5-mm-thick stainless-steel vacuum window, a 1-mm-thick, 4-cm-diameter lead synchrotron-radiation filter, and a 5-cm-thick, 8-cm-diameter lead collimator with an interchangeable aperture up to 2~cm in diameter. Thicker lead filters were available, but were shown in simulation to distort the energy spectrum. The final item in the beamline is the GSO crystal, a cylinder 6~cm in diameter and 15~cm (10.9~radiation lengths) long; Table~\ref{tab:gso_properties} lists the other crystal properties used in the MC. Figure~\ref{fig:posfirstEdep} histograms the locations where photons in the simulation first interact with matter. 

GEANT4 modeling of the gamma shower relies on the cross-section $\sigma$ for gamma conversion into an $(e^+, e^-)$ pair; for gamma energies between 1.5~MeV and 100~GeV, the parameterization is accurate to within 5\%, with a mean accuracy of 2.2\%~\cite{Geant_PRM}. Electromagnetic interactions~\cite{g4_em_physlists} are modeled based on the Livermore physics list; substituting the standard electromagnetic physics list or the Penelope physics list did not produce noticeable differences. Hadronic interactions are based on the QGSP\_BIC physics list; other packages, not designed for this energy range, gave a discrepancy of about 0.5\% in the average energy deposit. The combined non-linearity of the photomultiplier tube and front-end electronics was measured in situ~\cite{Friend:Pulser11}, and the resulting functional form is an input to the simulation. 

\begin{table}[tbdp]
\begin{center}
\begin{tabular}{ccc}
\hline
\textbf{Property} & \textbf{Value} & \textbf{Reference} \\
\hline
Density & $6.71$~g/cm$^3$ & \cite{Melcher:GSOscintillation1990} \\
Radiation length & $1.38$~cm & \cite{Tamagawa:ConnectedGSOProton_2006} \\
Attenuation length & $340$~cm & estimated from~\cite{Tamagawa:ConnectedGSOProton_2006} \\
Birks' constant & $5.25$~$\upmu$m/MeV & \cite{Avdeichikov:2000fk} \\
\end{tabular}
\end{center}
\caption
  {GSO:Ce (0.5~mol\%) properties used in simulation.}
\label{tab:gso_properties}
\end{table}

An optical extension to this basic simulation follows the path of each scintillation photon produced in the electromagnetic shower from an incident Compton-scattered photon. This package approximates the polished GSO surface as perfectly smooth, and includes the aluminum-foil detector wrapping (with a modeled reflectivity of 0.9) and the efficiency of the PMT photocathode. The output of the MC is the number of photoelectrons produced in the simulated photocathode. The resulting spectrum shows non-Gaussian smearing due to optical effects such as shower leakage from the crystal and the dependence of photon-collection efficiency on the initial interaction position. However, it takes 6000 times longer to generate a given number of events with the optical package than it does without the package, which is prohibitive for some simulation tasks.

Spectra simulated under Jefferson Lab conditions must be modified to take into account pileup, in which two or more pulses arrive and are integrated during the readout window for a single trigger. This correction is determined experimentally based on periodically triggered snapshots of the readout window, which give the random-event rate and the spectrum of energy deposited by random pulses during such a window. For spectra from electron-photon coincidence data (Section~\ref{sec:jlab}), it is sufficient to use this spectrum to add random, pileup pulse integrals to those of simulated Compton-scattered photons. For photon-arm singles data, however, the correction is complicated by the possibility of pileup between two Compton events; a naive pileup correction would count such a pair twice. This double-counting effect is canceled by a careful construction of the empirical pileup spectrum that is added to the MC: half from dedicated background measurements and half from Compton-scattering data~\cite{friend:2012fk}. For simulations of the primary running mode, in which signals are integrated over a long, untriggered window, no pileup correction is necessary.

\begin{figure}[] 
\begin{center} 
\includegraphics[width=\columnwidth]{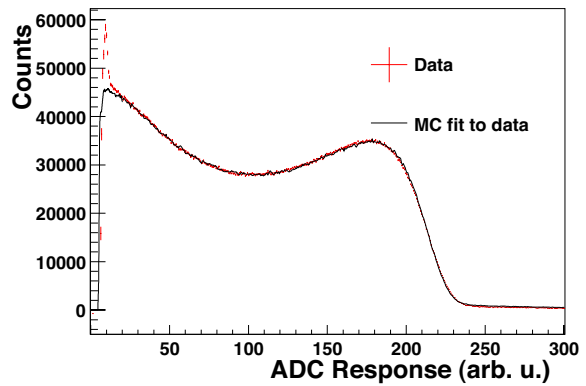}
\caption[Measured Compton Spectrum with MC Fit]{A Compton spectrum measured in Hall A, with the ADC response given in arbitrary units. The Compton edge in this configuration was at 204~MeV. A non-optical MC with 2.3\% Gaussian smearing is fit to the experimental data with only two free parameters -- a horizontal scale factor and a vertical scale factor -- with a $\chi^2/dof$ of 6.166.  The discrepancy at low photon energies is due to a shift in the effective trigger threshold caused by a rate-dependent gain shift in the timing filter amplifier. In the low-energy region, this threshold shift results in incorrect background subtraction; accordingly, this region is excluded from the fit.
\label{fig:comptSpect}} 
\end{center}
\end{figure}

A 2.3\% Gaussian smearing function must be folded into the basic, non-optical MC for a satisfactory match to the experimental data, as shown in Fig.~\ref{fig:comptSpect}. By contrast, the optical MC requires only 1.5\% Gaussian smearing; that is, a 1.5\% smearing is not accounted for by modeled optical effects. Timing jitter of the photon pulse within the trigger window, which removes a varying amount of the tail from the signal integration, accounts for about 0.7\% smearing. Other candidates include unmodeled optical properties such as an imperfect surface polish, and unmeasured non-linearities in the system response, which would have a non-negligible effect on the analyzing power of the polarimeter. The required smearing factors were determined by a comparison of fits that assumed various levels of smearing; they were not parameters in the fits. Photon-singles data from Hall A of Jefferson Lab and from HI$\upgamma$S (Section~\ref{sec:higs}) were used for these comparisons. At high data rates, pileup corrections are necessary for both optical and non-optical models. 

\section{Tests with Tagged Photons at Jefferson Lab} \label{sec:jlab}

\begin{figure}[] 
\begin{center}
\subfigure[]{\label{fig:strip10}\includegraphics[width=\columnwidth]{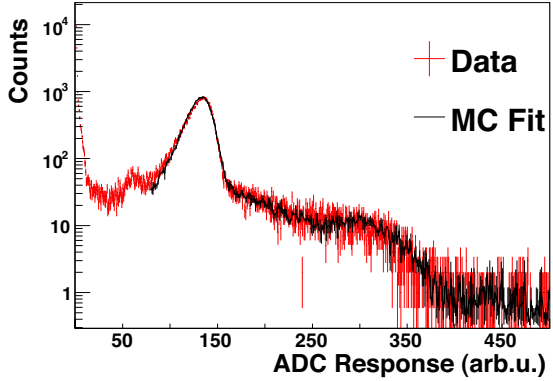}}
\subfigure[]{\label{fig:strip36}\includegraphics[width=\columnwidth]{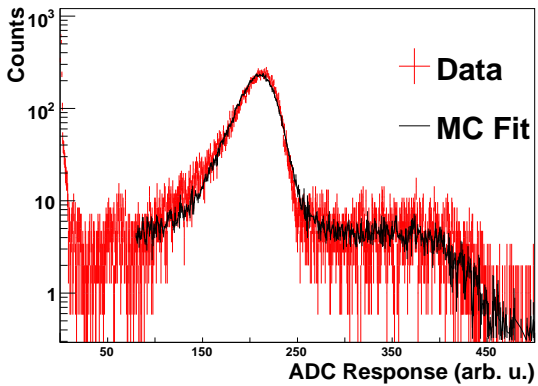}}
\caption{Fits of the non-optical MC, including background, to tagged photon spectra from \subref{fig:strip10}~strip~10 ($125.1-127.0$-MeV photons, $\chi^2/dof=2.13$) and \subref{fig:strip36}~strip~36 ($203.1-204.9$-MeV photons, $\chi^2/dof=1.76$).  
\label{fig:EDETstrips} } 
\end{center} 
\end{figure}

\begin{figure}[] 
\begin{center} 
\includegraphics[width=\columnwidth]{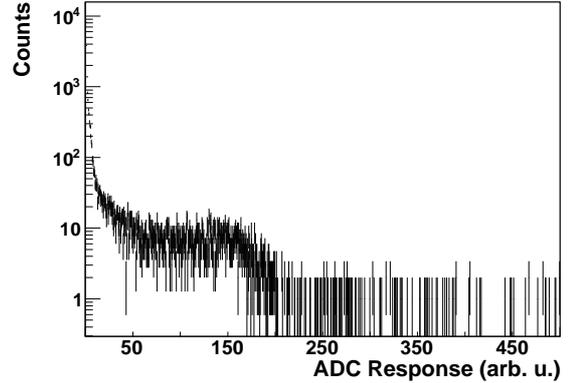}
\caption[Empirical pileup spectrum with a 3.4-GeV electron beam]{Measured spectrum of pileup pulses in tagged-photon data, used to correct MC spectra.
\label{fig:pileuplog}} 
\end{center} 
\end{figure}

Further tests of the GSO calorimeter in Hall A were performed with photons tagged by the electron detector located above and downstream of the third dipole of the Compton polarimeter. Each of this detector's four planes consists of 192 horizontal strips with a 240-$\upmu$m pitch. When a scattered electron strikes the detector, the vertical position of the strip gives its deflection angle in the third dipole and therefore its momentum, which in turn determines the energy of the associated scattered photon. The electron detector can thus tag the energies of coincident photons seen in the GSO crystal. For these data, taken with an initial electron-beam energy of 3.4~GeV, the maximum backscattered photon energy, or Compton edge, was 204~MeV. Therefore, only the bottom 37~strips saw Compton-scattered electrons.

Fig.~\ref{fig:EDETstrips} shows measured and modeled spectra of photons tagged by two typical strips. GSO readout was triggered by the electron detector. Due to computational constraints, the MC is non-optical, but it does include simulated background events and pileup effects; Fig.~\ref{fig:pileuplog} shows the measured pileup spectrum used for the MC.  Each fit has three free parameters: a vertical scale factor, a horizontal scale factor, and a factor setting the amount of included background. The latter varies from strip to strip because electron-detector noise, which is slightly different for each strip, is a significant source of background. An enhancement of the background at 60~arb.~u.~(Fig.~\ref{fig:EDETstrips}) appears in each energy bin; it is believed to be the result of Compton-scattered electrons near the kinematic limit, which impinge on the electron-detector shielding and produce hits in every strip. The associated Compton-scattered photons then appear with all energy tags. These photons actually have a narrow, fixed energy range centered at 60~arb.~u., since only electrons in a certain energy range strike the shielding material. This complex background source was not implemented in the MC, as it has no significance in non-tagged data sets.

MC fits for all 37~strips, performed in the range above the background from conversion in the shielding, are given a common horizontal offset and 5\% Gaussian smearing. The larger smearing requirement cannot be understood solely from photon-detector data. Neither the dispersion nor the fringe fields in the third dipole are well known; these likely contribute a significant portion of the broadening, along with an additional contribution from the poorly understood electron-detector noise.

\section{Tests with $20-40$-MeV Photons at HI$\upgamma$S}
\label{sec:higs}

HI$\upgamma$S uses Compton scattering to produce a gamma beam, the energy of which ranged from 2 to 65~MeV at the time of our tests.  At HI$\upgamma$S, electron bunches that circulate in a storage ring are sent through magnetic undulators, causing the electrons to produce free-electron laser (FEL) light, which is stored in an optical cavity.  The FEL photons are then allowed to collide with electrons in the storage ring; the Compton-backscattered photons travel through a manually adjustable collimator. The result is a nearly monoenergetic gamma beam in the experimental hall, approximately 60~m downstream of the Compton interaction point.

\begin{figure}[] 
\begin{center} 
\subfigure[]{\label{fig:HIGSfit20}\includegraphics[width=\columnwidth]{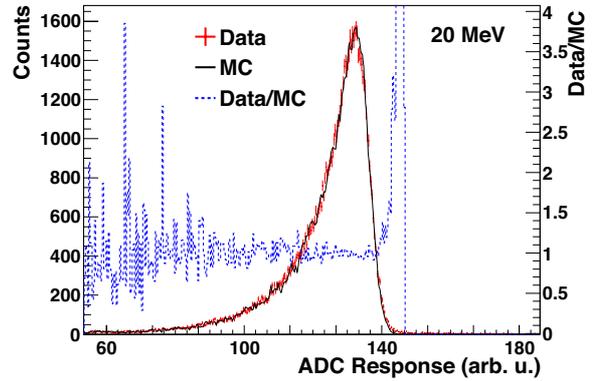}}
\subfigure[]{\label{fig:HIGSfit3energies}\includegraphics[width=\columnwidth]{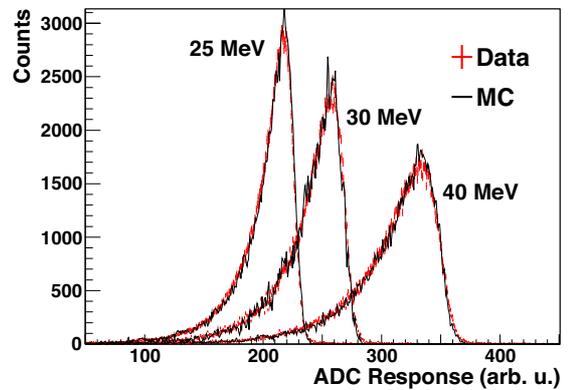}}
\subfigure[]{\label{fig:HIGSfit40}\includegraphics[width=\columnwidth]{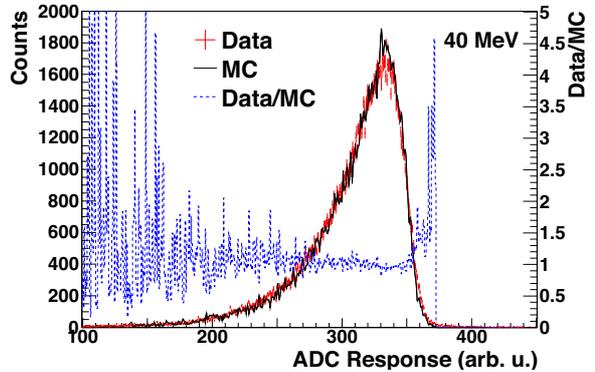}}
\caption[Measured and modeled spectra from tests at HI$\upgamma$S]{Simulated spectra fit with two free parameters (horizontal and vertical scale factors) to experimental data from the HI$\upgamma$S facility, for incident photons at~\subref{fig:HIGSfit20} 20~MeV ($\chi^2/dof=3.26$, plotted with the ratio of measured counts to simulated counts (dashed line)),~\subref{fig:HIGSfit3energies} 25, 30, and 40~MeV, with a combined fit for all three energies ($\chi^2/dof=10.20$), and~\subref{fig:HIGSfit40} the measured and simulated spectra for 40-MeV photons, with the ratio of measured counts to simulated counts (dashed line). Gain shifts prevent combining the fits to the 20-MeV and 22-MeV (not shown) spectra with the higher-energy spectra.
\label{fig:HIGSfitconst}} 
\end{center} 
\end{figure}

\begin{table}[tbdp]
\begin{center}
\begin{tabular}{ccc}
\hline
\textbf{Photon Energy} & \textbf{Photon Energy}  & \textbf{$E_e$} \\
\textbf{(MeV)} & \textbf{FWHM (MeV)} & \textbf{(MeV)} \\
\hline
20 & 1.0 & 538 \\
22 & 1.2 & 565 \\
25 & 1.5 & 603\\
30 & 2.1 & 662 \\
40 & 3.7 & 767 \\
\end{tabular}
\end{center}
\caption
  {HI$\upgamma$S photon beam properties, and estimated electron-beam energy $E_e$, during GSO tests. The FEL lasing wavelength was 265~nm.}
\label{tab:higs_beam}
\end{table}

A series of HI$\upgamma$S-beam test runs at five different photon energies was taken using the GSO calorimeter; Table~\ref{tab:higs_beam} shows the run configurations. Gain shifts after the 20-MeV and 22-MeV data points, likely due to difficulties with the high-voltage supply for the photomultiplier tube, prevent the analysis of all five energies together. Figure~\ref{fig:HIGSfitconst} shows the measured spectra at four energy settings, as well as the ratio of measured to modeled counts for 20-MeV and 40-MeV photons. An optical MC of the HI$\upgamma$S configuration, including non-linearity, was fit to each spectrum with two free parameters: horizontal and vertical scale factors. Background and pileup effects were negligible at all energies except 22~MeV, and were not included in the MC.

Because the energy spread of the HI\(\upgamma\)S beam is well known, fits to these data constrain the minimum amount of additional smearing that must be included to bring the MC into agreement with the data. The corresponding 1.5\% smearing is included in the simulations shown in Fig.~\ref{fig:HIGSfitconst}. 

\section{Conclusion} \label{sec:conclusion}

\begin{figure}[]
\begin{center}
\includegraphics[width=\columnwidth]{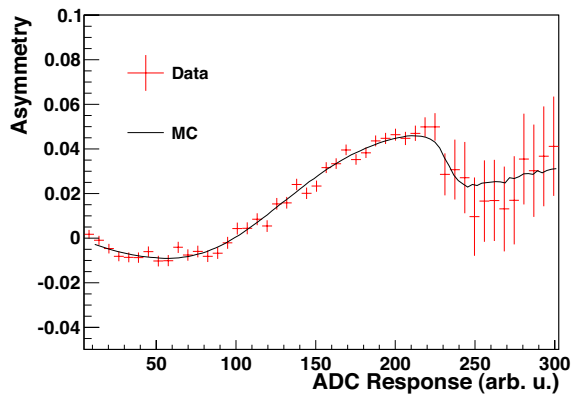} 
\caption[]{The measured Compton asymmetry plotted as a function of ADC response to deposited photon energy.  The experimental data are compared to GEANT4 MC data with no adjustable parameters. The Compton edge in this configuration was at 204~MeV. 
\label{fig:asymSpect}} 
\end{center}
\end{figure}

We have demonstrated agreement at the few-percent level between the MC of this GSO:Ce crystal and photon spectra measured over a range of rates and energies. The ultimate purpose of the MC, however, was to unfold the detector response from Compton-photon polarimetry data in Hall A. The asymmetry in Compton-scattering cross-sections for opposite photon-and-electron spin configurations is proportional to the polarizations of both incident beams and to the energy-weighted analyzing power of the apparatus~\cite{Friend:ComptonUpgrade2011}; the MC is a vital input to understanding this analyzing power and its uncertainty. 

With the aid of the MC, we determined that the analyzing power was insensitive to a misalignment of up to 5~mm between the scattered-photon beam and the collimator, and to a misalignment of up to 10~mm between the scattered-photon beam and the GSO crystal~\cite{friend:2012fk}. We quantified the effect of the 1-MeV uncertainty in determining the electron-beam energy in Hall A: a shift of that size in a 3.4-GeV beam energy changes the analyzing power by 0.1\%~\cite{friend:2012fk}. A Gaussian smearing factor is required to produce the fits shown here; it does not directly affect the analyzing power, but it does place a limit on the unknown non-linearity in the system response. We conclude that this unknown response contributes a maximum 0.3\% systematic error to the analyzing power~\cite{friend:2012fk}. 

Figure~\ref{fig:asymSpect} shows the Compton asymmetry as a function of the energy deposited in the GSO by Compton-scattered photons, as measured in counting mode. The non-optical MC result is plotted on the same axes with no adjustable parameters. The horizontal scale was determined from a fit to the full photon energy spectrum (Fig.~\ref{fig:comptSpect}); the vertical scale is set by the incident photon polarization and by the electron beam polarization, which was determined from energy-weighted integration-mode data~\cite{Friend:ComptonUpgrade2011} and confirmed by M{\o}ller polarimetry measurements~\cite{happex3:2012fk}. 

The analyzing power for the GSO calorimeter is now known to within 0.33\%, dominated by our imperfect knowledge of the system non-linearity~\cite{Friend:ComptonUpgrade2011}. Existing GEANT4 physics modules allow simulation of the GSO:Ce response to gammas in the $20-200$-MeV energy range with satisfactory precision.

We gratefully acknowledge the assistance of staff and scientists at HI$\upgamma$S, especially Mohammad Ahmed, Sean Stave, and Ying Wu, and at Jefferson Lab, especially the Hall A Compton laser working group. This work was supported by DOE grant DE-FG02-87ER40315. Jefferson Lab is operated by the Jefferson Science Associates, LLC, under DOE grant DE-AC05-060R23177.  HI$\upgamma$S is operated jointly by the Triangle Universities Nuclear Laboratory and by the Duke Free Electron Laser Laboratory.


\bibliographystyle{model1a-num-names} \bibliography{BibComptonGSO2011}


\end{document}